\documentclass[10pt,aps,prd,fleqn,notitlepage,superscriptaddress,nofootinbib,preprintnumbers]{revtex4-1}
\pdfoutput=1
\usepackage{amsmath,amssymb,graphicx,xspace,subfigure,relsize,todonotes}
\newcommand{\UNNLOPS}{\protect\scalebox{0.9}{UN$^2$LOPS}\xspace}
\newcommand{\UNLOPS}{\protect\scalebox{0.9}{UNLOPS}\xspace}
\newcommand{\MENLOPS}{\protect\scalebox{0.9}{MENLOPS}\xspace}

\newcommand{\POWHEG}{\protect\scalebox{0.9}{POWHEG}\xspace}
\newcommand{\MCatNLO}{\protect\scalebox{0.9}{MC@NLO}\xspace}
\newcommand{\MINLO}{\protect\scalebox{0.9}{MINLO}\xspace}

\newcommand{\HNNLO}{\protect\scalebox{0.9}{HNNLO}\xspace}
\newcommand{\MCFM}{\protect\scalebox{0.9}{MCFM}\xspace}
\newcommand{\abr}[1]{\langle #1\rangle}
\newcommand{\mc}[1]{\mathcal{#1}}
\newcommand{\mr}[1]{\mathrm{#1}}

\newcommand{\done}{{\rm d}}
\newcommand{\order}{\mathcal{O}}
\newcommand{\bs}{\!\!\!\!\!\!}
\usepackage[pdfborder={0 0 0}]{hyperref}
\hypersetup{
  pdfauthor={Stefan Hoeche, Ye Li, Stefan Prestel},
  pdftitle={Higgs-boson production through gluon fusion at NNLO QCD with parton showers}
}

\begin{document}
\preprint{SLAC-PUB 16011}
\preprint{DESY 14-119}
\preprint{MCNET 14-14}

\title{Higgs-boson production through gluon fusion at NNLO QCD with parton showers}

\author{Stefan~H{\"o}che}
\affiliation{SLAC National Accelerator Laboratory, Menlo Park, CA 94025, USA}
\author{Ye~Li}
\affiliation{SLAC National Accelerator Laboratory, Menlo Park, CA 94025, USA}
\author{Stefan Prestel}
\affiliation{Deutsches Elektronen-Synchrotron, DESY, 22603 Hamburg, Germany}

\begin{abstract}
We discuss how the \UNNLOPS scheme for matching NNLO calculations
to parton showers can be applied to processes with large higher-order 
perturbative QCD corrections. We focus on Higgs-boson production 
through gluon fusion as an example. We also present an NNLO fixed-order
event generator for this reaction.
\end{abstract}

\maketitle

\section{Introduction}
The high precision of experimental measurements at the Large Hadron Collider (LHC)
requires equally precise theoretical calculations for the Standard Model
processes of interest, such as Higgs-boson production~\cite{Dittmaier:2011ti,
  *Dittmaier:2012vm,*Heinemeyer:2013tqa}. Experimental analyses often 
impose intricate kinematical cuts on the final-state phase space, thus
calling for fully differential predictions of the cross section. At the same
time the resummation of large logarithmic corrections is important, especially
in order to describe QCD radiative corrections to the hard process.

These requirements have inspired many new techniques to simulate the structure
of hadron collider events with unprecedented accuracy~\cite{Buckley:2011ms}.
Among them are the pioneering matching methods \MCatNLO~\cite{Frixione:2002ik}
and \POWHEG~\cite{Nason:2004rx,*Frixione:2007vw}, which allowed, for the first time,
to combine next-to-leading order (NLO) QCD calculations with parton showers
by means of a modified subtraction scheme. Even higher precision is needed
for Higgs physics, as the dominant production mode through gluon fusion 
suffers from large perturbative corrections. Next-to-next-to leading order (NNLO)
accurate fixed-order predictions~\cite{Anastasiou:2002yz,*Harlander:2002wh,*Anastasiou:2005qj,
  Catani:2007vq,Grazzini:2008tf,*Catani:2008me} are 
routinely employed, NNLO accurate predictions for Higgs plus jet production have been 
obtained~\cite{Boughezal:2013uia} and N$^3$LO accurate results are 
actively pursued~\cite{Anastasiou:2013mca,*Kilgore:2013gba,*Anastasiou:2014vaa,*Ball:2013bra,
  *Li:2013lsa,*Duhr:2013msa,*Li:2014bfa}.
Mixed QCD and electroweak two-\-loop corrections have been estimated 
assuming complete factorization~\cite{Actis:2008ug} and later evaluated
in an effective theory approach~\cite{Anastasiou:2008tj}.
Resummed predictions have been made at NNLO+NNLL accuracy~\cite{deFlorian:2009hc,*deFlorian:2012mx,Becher:2012yn},
and jet vetoed cross sections, particularly relevant for Higgs boson decay
channels involving $W$ bosons have been in the focus of interest recently~\cite{
  Banfi:2012yh,*Banfi:2012jm,*Bonvini:2014joa,*Stewart:2013faa,
  *Becher:2012qa,*Becher:2013xia}. The matching of NNLO
fixed-order results to parton showers using the \MINLO method was presented
in~\cite{Hamilton:2012rf,*Hamilton:2013fea}.
 
Making the most precise fixed-order predictions available in the framework of
general purpose event generators is a challenging task. Three different proposals
exist for matching NNLO calculations to parton showers~\cite{
  Lavesson:2008ah,Hamilton:2012rf,*Hamilton:2013fea,Alioli:2013hqa}, but only two of them 
were implemented so far~\cite{Lavesson:2008ah,Hamilton:2012rf,*Hamilton:2013fea,Hoeche:2014aia,Karlberg:2014qua}.
The aim of this publication is to discuss one of them, the \UNNLOPS matching method, for 
Higgs boson production at hadron colliders. We also present an independent,
fully differential NNLO fixed-order calculation of Higgs-boson production
using the $q_T$-cutoff technique.

This article is organized as follows: Section~\ref{sec:unlops} reviews the \UNNLOPS
matching method. Section~\ref{sec:nnlo} discusses the implementation of the NNLO
calculation. Alterations of the \UNNLOPS proposal~\cite{Hoeche:2014aia},
needed for the matching in Higgs production are introduced in Sec.~\ref{sec:higgs}.
Section~\ref{sec:results} presents our results and Sec.~\ref{sec:conclusions} gives
an outlook.

\section{UN$^2$LOPS matching}
\label{sec:unlops}
To set the stage, we recall the refined \UNNLOPS method~\cite{Lonnblad:2012ix} introduced
in~\cite{Hoeche:2014aia}. We assume \MCatNLO matched predictions~\cite{Frixione:2002ik},
for Higgs-boson plus zero and one jets, which are to be merged. Any infrared safe observable
$O$ is computed in these simulations as
\begin{equation}\label{eq:mcatnlo}
  \abr{O}_n=\int\done\Phi_n\tilde{\mr{B}}_n(\Phi_n)\bar{\mc{F}}_n(t(\Phi_n),O)
  +\int\done\Phi_{n+1}\mr{H}_{n}(\Phi_{n+1})\mc{F}_{n+1}(t(\Phi_{n+1}),O)\;,
\end{equation}
where $\done\Phi_n$ denotes the differential $n$-particle phase space element, including a 
convolution with the PDFs. We use the following notation for the NLO-weighted Born 
cross section, $\tilde{\mr{B}}$, and the hard remainder function, $\mr{H}$:
\begin{equation}\label{eq:mcatnlo_bbar_h}
  \begin{split}
    \tilde{\mr{B}}_n(\Phi_n)=&\;\mr{B}_n(\Phi_n)+\tilde{\mr{V}}_n(\Phi_n)+\mr{I}_n(\Phi_n)
    +\int\done\hat{\Phi}_1\Big[\mr{S}_n(\Phi_n,\hat{\Phi}_1)\,\Theta(t_n(\Phi_n)-t_{n+1}(\hat{\Phi}_1))-\mr{D}_n(\Phi_n,\hat{\Phi}_1)\Big]\\
    \mr{H}_{n}(\Phi_{n+1})=&\;\mr{B}_{n+1}(\Phi_{n+1})-\mr{D}_{n}(\Phi_{n+1})\Theta(t_n(\Phi_n)-t_{n+1}(\Phi_{n+1}))\;.
  \end{split}
\end{equation}
The functions $\mr{B}_n$, $\tilde{\mr{V}}_n$ and $\mr{S}_n$ represent the Born, virtual,
and real subtraction contribution to the $n$-jet NLO calculation, while $\mr{I}_n$ stands
for the integrated subtraction terms~\cite{Hoeche:2011fd}. $\done\hat{\Phi}_1$ represents 
the one-emission differential phase-space element, which factorizes as 
$\done\hat{\Phi}_1=\done t\,\done z\,\done\phi/(2\pi)\,J(t,z)$, with $J(t,z)$ a Jacobian factor. 

The generating functional of the parton shower, $\mc{F}_n(t,O)$, is defined using 
the parton-shower evolution kernels, $\mr{K}_n$.
\begin{equation}\label{eq:gen_func_ps}
  \mc{F}_n(t,O)=\Pi_n(t_c,t)\,O(\Phi_n)
  +\int_{t_c}^t\done\hat{\Phi}_1\,\mr{K}_n(\hat{\Phi}_1)\,
  \Pi_n(\hat{t},t)\,\mc{F}_{n+1}(\hat{t},O)\;.
\end{equation}
The no-branching probability, $\Pi_n$, follows from the unitarity condition $\mc{F}_n(t,1)=1$.
We use the abbreviations $t_n=t(\Phi_n)$ and $\hat{t}=t(\hat{\Phi}_1)$ to denote the
evolution scales for the $n$-particle process and the additional emission generated
in the integration over $\done\hat{\Phi}_1$. Similarly, we define a generating
functional for the \MCatNLO.
\begin{equation}\label{eq:gen_func_mcatnlo}
  \bar{\mc{F}}_n(t,O)=\tilde{\Pi}_n(t_c,t)\,O(\Phi_n)
  +\int_{t_c}^t\done\hat{\Phi}_1\,\frac{\mr{D}_n(\Phi_n,\hat{\Phi}_1)}{\mr{B}_n(\Phi_n)}\,
  \tilde{\Pi}_n(\hat{t},t)\,\mc{F}_{n+1}(\hat{t},O)\;.
\end{equation}
We restrict real corrections in the zero-jet \MCatNLO to transverse momenta $q_T<q_{T,\rm cut}$
(denoted by by $\mr{H}_1^{q_{T,\rm cut}}$), and we choose $q_{T,\rm cut}$ such that it falls below 
the parton-shower cutoff, $t_c$. At the same time the one-jet calculation is regularized by requiring
Higgs-boson transverse momenta larger than $q_{T,\rm cut}$. This implies that resolved real corrections
are generated solely by the one-jet \MCatNLO. In order to reproduce the logarithmic structure of the 
parton-shower prediction, the emission terms must be weighted by the all-order resummed virtual 
and unresolved corrections~\cite{Amati:1980ch}, which are obtained in form of a no-branching probability
computed by the parton shower. At the same time, coupling renormalization and some higher-logarithmic 
corrections~\cite{Dokshitzer:1978qu,*Amati:1978by,
 *Ellis:1978sf,*Libby:1978ig,*Mueller:1978xu,*Dokshitzer:1978hw,Catani:1990rr} are accounted for
by reweighting with the appropriate ratio of coupling constants.
This reweighting introduces $\mathcal{O}(\alpha_s)$ terms that impair the NLO accuracy. 
They are subtracted by the first-order expansion of the weight factor, and by the first-order 
expansion of the no-branching probability, $\Pi_0^{(1)}(t_1,\mu_Q^2)$~\cite{Hoeche:2014aia}. 
\begin{equation}
  \begin{split}
    \abr{O}_1=&\;\int_{q_{T,\rm cut}}\bs\done\Phi_1\,\Big[\,\mr{B}_1(\Phi_1)\,\Pi_0(t_1,\mu_Q^2)
    \Big(w_1(\Phi_1)+w_1^{(1)}(\Phi_1)+\Pi_0^{(1)}(t_1,\mu_Q^2)\Big)
    +\tilde{\mr{B}}^{\rm R}_1(\Phi_1)\,\Pi_0(t_1,\mu_Q^2)\,\Big]\,\bar{\mc{F}}_1(t_1,O)\\
    &+\int_{q_{T,\rm cut}}\bs\done\Phi_2\,\Big[\,\mr{H}^{\rm R}_1(\Phi_1)\,\Pi_0(t_1,\mu_Q^2)
    +\mr{H}^{\rm E}_1(\Phi_1)\,\Big]\,\mc{F}_2(t_2,O)\;.
  \end{split}
\end{equation}
We have defined $\tilde{\mr{B}}^{\mr R}=\tilde{\mr{B}}-\mr{B}$ and the regular and exceptional
parts of the real corrections, $\mr{H}^{\rm R}$ and $\mr{H}^{\rm E}$. They correspond to two-parton
configurations with and without a parton-shower equivalent, respectively~\cite{Hoeche:2014aia}.
$\mu_Q^2$ defines the resummation scale. The weight factor $w_1$ is given as
\begin{equation}\label{eq:unlops_weight}
  w_1(\Phi_1)=\frac{\alpha_s(b\,t_1)}{\alpha_s(\mu_R^2)}\,
  \frac{f_a(x_a,t_1)}{f_a(x_a,\mu_F^2)}
  \frac{f_{a'}(x_{a'},\mu_F^2)}{f_{a'}(x_{a'},t_1)}
  \qquad\text{where}\qquad
  \beta_0\ln\frac{1}{b}=\bigg(\frac{67}{18}-\frac{\pi^2}{6}\bigg)C_A-\frac{10}{9}\,T_R\,n_f\;,
\end{equation}
and where $f_a(x_a)$ and $f_{a'}(x_{a'})$ denote the PDFs associated with the external
and intermediate parton (in the sense of a parton-shower history).
The scale factor $b$ includes effects of the 2-loop cusp anomalous dimension 
in the parton shower~\cite{Kodaira:1982az,*Catani:1989ne,*Catani:1988vd,*Catani:1992ua,Catani:1990rr}.

The restricted zero-jet calculation and the one-jet \MCatNLO result above $q_{T,\rm cut}$
do not overlap. We can thus replace the zero-jet \MCatNLO by a full $q_T$-vetoed NNLO calculation,
using the cutoff method~\cite{Gao:2012ja}, and complete the cross section using the one-jet 
\MCatNLO~\cite{Hoeche:2014aia}: Each event removed from the one-jet bin by means of the emission
probability $1-\Pi_0(t,\mu_Q^2)$ is added to the zero-jet bin. This unitarization procedure is equivalent
to a parton-shower resummation of the jet veto cross section at $q_{T,\rm cut}$. It gives the
\UNNLOPS matching formula
\begin{equation}\label{eq:nnlo_ps}
  \begin{split}
    &\abr{O}^\mr{(UN^2LOPS)}=\;
    \int\done\Phi_0\,\bar{\bar{\mr{B}}}_0^{q_{T,\mr{cut}}}(\Phi_0)\,O(\Phi_0)\\
    &\quad+\int_{q_{T,\mr{cut}}}\bs\done\Phi_1\,
    \Big[1-\Pi_0(t_1,\mu_Q^2)\,
      \Big(w_1(\Phi_1)+w_1^{(1)}(\Phi_1)+\Pi_0^\mr{(1)}(t_1,\mu_Q^2)\Big)\Big]\,
    \mr{B}_1(\Phi_1)\,O(\Phi_0)\\
    &\quad+\int_{q_{T,\mr{cut}}}\bs\done\Phi_1\,
    \Pi_0(t_1,\mu_Q^2)\Big(w_1(\Phi_1)+w_1^{(1)}(\Phi_1)+\Pi_0^\mr{(1)}(t_1,\mu_Q^2)\Big)
    \,\mr{B}_1(\Phi_1)\,\bar{\mc{F}}_1(t_1,O)\\
    &\quad+\int_{q_{T,\mr{cut}}}\bs\done\Phi_1\,
      \Big[1-\Pi_0(t_1,\mu_Q^2)\Big]\,\tilde{\mr{B}}_1^{\rm{R}}(\Phi_1)\,O(\Phi_0)
    +\int_{q_{T,\mr{cut}}}\bs\done\Phi_1
    \Pi_0(t_1,\mu_Q^2)\,\tilde{\mr{B}}_1^{\rm{R}}(\Phi_1)\,\bar{\mc{F}}_1(t_1,O)\\
    &\quad+\int_{q_{T,\mr{cut}}}\bs\done\Phi_2\,
      \Big[1-\Pi_0(t_1,\mu_Q^2)\Big]\,\mr{H}_1^{\mr{R}}(\Phi_2)\,O(\Phi_0)
    +\int_{q_{T,\mr{cut}}}\bs\done\Phi_2\,
      \Pi_0(t_1,\mu_Q^2)\,\mr{H}_1^{\mr{R}}(\Phi_2)\,\mc{F}_2(t_2,O)\\
    &\quad+\int_{q_{T,\mr{cut}}}\bs\done\Phi_2\,
      \,\mr{H}_1^{\mr{E}}(\Phi_2)\,\mc{F}_2(t_2,O)\;,
  \end{split}
\end{equation}
where $\bar{\bar{\mr{B}}}_0^{q_{T,\mr{cut}}}$ represents the $q_T$-vetoed NNLO
cross section, differential in the Born phase space. By construction the inclusive cross
section exactly reproduces the NNLO result.

\section{The $\mathbf{q_T}$-vetoed NNLO calculation}
\label{sec:nnlo}
In the dominant production mode at hadron colliders, the Higgs boson couples to two gluons 
via heavy quark loops. The full top and bottom quark mass dependent gluon fusion cross section 
is known to NLO only~\cite{Spira:1996if,*Anastasiou:2009kn}.
It is more convenient to work in an effective theory (HEFT), where the heavy top quark is 
integrated out~\cite{Ellis:1975ap,*Wilczek:1977zn,*Shifman:1979eb,*Ellis:1979jy}.
The gluon fusion process is then described by the effective Lagrangian,
\begin{equation}\label{eq:heft_operator}
  \mathcal{L}_{\rm eff} = -\frac{\alpha_s}{4 v} \frac{c_g}{3 \pi} H G^{a}_{\mu\nu} G_{a}^{\mu\nu},
\end{equation}
where $v$ is the Higgs vacuum expectation value, and $c_g=1+\mathcal{O}(\alpha_s)$
is the Wilson coefficient. Quark mass effects can be incorporated a posteriori. 
The matching to the Higgs effective theory leads to a factorized 
form of the hard function. We write it schematically as,
\begin{equation}\label{eq:hard_function}
  H(Q^2,\mu^2) = |c_g|^2 = \sum_{n=0} \left( \frac{\alpha_s(\mu^2)}{4 \pi} \right)^n h^{(n)}(Q^2,\mu^2)\;.
\end{equation}
The hard function can be improved by including an addition overall factor of the full 
top mass dependent LO gluon fusion cross section, normalized by the HEFT LO cross section.
This is a very good approximation at NNLO~\cite{Pak:2009dg,Harlander:2009mq,*Harlander:2009my}.
The hard function is applied to the NNLO results of effective theory order by order in $\alpha_s$,
i.e.\ $h^{(2)}$ only multiplies the LO HEFT cross section, $h^{(1)}$ multiplies the NLO HEFT 
cross section, and $h^{(0)}$ multiplies the NNLO HEFT cross section. A similar scheme can be used 
in the matched calculation. However, we can also multiply the hard function as an overall factor 
to the HEFT calculation, leading to a second possible matching scheme, which will be discussed 
in Sec.~\ref{sec:higgs}.

The NNLO Higgs production in the Higgs effective theory implemented in Sherpa is based on 
the $q_T$ subtraction method~\cite{Catani:2007vq,Catani:2009sm}. It separates the two-loop
NNLO calculation into a zero-$q_T$ bin, leaving the phase space with finite $q_T$ to the
NLO calculation. This matches well with the general structure of \UNNLOPS, introduced in Sec.~\ref{sec:unlops}.
The dependence on the $q_T$ cutoff, used to define the zero-$q_T$ bin, cancels between contributions
from the two phase space regions. Given a small enough $q_T$ cutoff, the zero-$q_T$ bin can be mapped
onto the Born phase space, as all soft and collinear radiation is integrated over. 
For Higgs or Drell Yan processes, where there is no final state colored particle at Born level,
the $q_T$ cut roughly corresponds to the parton shower cutoff scale. In addition, 
the zero-$q_T$ bin follows a simple factorization formula, which generates very compact results
for $\bar{\bar{\mr{B}}}_0^{q_{T,\mr{cut}}}$ that can easily be implemented numerically. 
The remainder is computed as Higgs-boson plus one-jet production at NLO, using \cite{Ravindran:2002dc}
for the virtual matrix elements and the Catani-Seymour subtraction method for regularizing real radiative
corrections~\cite{Catani:1996vz}.

As a consequence of the factorization, the zero-$q_T$ bin contribution can be written
as a differential K factor to the Born level cross section
\begin{equation}\label{eq:smallqtfact}
  \frac{ \int_0^{q_{T,\rm cut}} \done q_T \int_{x_i}^{1} 
    \done \xi_i \int_{x_j}^{1} \done \xi_j \; C_{ij \to gg}(q_{T},\frac{x_i}{\xi_i},\frac{x_j}{\xi_j},\mu)
    f_i(\xi_i,\mu_F^2) f_j(\xi_j,\mu_F^2) } {f_g(x_i,\mu_F^2) f_g(x_j,\mu_F^2) }\;,
\end{equation}
where $C_{ij \to gg}$ is the hard collinear coefficient, and $f_i(x_i,\mu_F^2)$ refers to the PDF.
The factorization formula describes the vetoed Higgs NNLO cross section up to power corrections 
in the cutoff, $q_{T,\rm cut}$.

The hard collinear coefficient is derived using the framework developed in~\cite{Becher:2012yn}
and using recent two loop results given in~\cite{Gehrmann:2014yya,*Gehrmann:2014uaa}. The results
have been verified against those presented in~\cite{Catani:2011kr}. In the framework of Sherpa,
our implementation is fully differential in phase space, which allows to generate events at the
parton level. Additionally, Higgs-boson decays to arbitrary final states can be included.  

\section{UN$^2$LOPS in Higgs-boson production}
\label{sec:higgs}
Higgs-boson production via gluon fusion suffers from large perturbative corrections,
both to the inclusive cross section and to the shape of distributions like 
the transverse momentum of the Higgs boson. This necessitates a careful treatment
of higher-order effects in the matching to parton showers. In processes like Drell-Yan 
lepton pair production -- where perturbative corrections to the transverse momentum 
distribution are small -- different matching schemes will lead to similar results
in the sense that any possible difference cannot be resolved experimentally. 
The situation is just the opposite in Higgs boson production, which was pointed out 
in several comparisons of NLO matching methods~\cite{Alioli:2008tz,Hoeche:2011fd,Nason:2012pr}.
We discuss the problem in this section, and we propose two different strategies to
tackle processes with large higher-order corrections.

Equation~\eqref{eq:nnlo_ps} proposes to subtract the no-branching probabilities, $\Pi_0$,
only when they multiply the leading-order part, $\mr{B}_1$, of the one-jet \MCatNLO result.
This is a minimal approach. The expansion of the no-branching probabilities in powers 
of the strong coupling generates terms of $\order(\alpha_s^3)$ when multiplied by 
$\tilde{\mr{B}}_1$ or $\mr{H}_1$, which is beyond the required NNLO accuracy of \UNNLOPS.
It is thus acceptable to also subtract these no-branching probabilities. A factorization
of the subtractions of $w_1$ and $\Pi_0$ multiplying the $\mr{B}_1$ term also only 
amounts to changing orders beyond the formal accuracy:
\begin{equation}
  \begin{split}
\Pi_0(t_1,\mu_Q^2)\left(w_1(\Phi_1)+w_1^{(1)}(\Phi_1)+\Pi_0^{(1)}(t_1,\mu_Q^2)\right)
\rightarrow
\Pi_0(t_1,\mu_Q^2)\left(1+\Pi_0^{(1)}(t_1,\mu_Q^2)\right)
\left(w_1(\Phi_1)+w_1^{(1)}(\Phi_1)\right)\,\mr{B}_1(\Phi_1)
  \end{split}
\end{equation}
Finally, evaluating PDFs and $\alpha_s$ in $\Pi_0^{(1)}$ with running scales
is legitimate. We will redefine $\Pi_0^{(1)}$ in this way below.
After these changes, the revised \UNNLOPS matching formula reads
\begin{equation}\label{eq:nnlo_ps_revised}
  \begin{split}
    &\abr{O}^\mr{(UN^2LOPS)}=\;
    \int\done\Phi_0\,\bar{\bar{\mr{B}}}_0^{q_{T,\mr{cut}}}(\Phi_0)\,O(\Phi_0)\\
    &\quad+\int_{q_{T,\mr{cut}}}\bs\done\Phi_1\,
    \Big[1-\Pi_0(t_1,\mu_Q^2)\Big(1+\Pi_0^\mr{(1)}(t_1,\mu_Q^2)\Big)\,
      \Big(w_1(\Phi_1)+w_1^{(1)}(\Phi_1)\Big)\Big]\,
    \mr{B}_1(\Phi_1)\,O(\Phi_0)\\
    &\quad+\int_{q_{T,\mr{cut}}}\bs\done\Phi_1\,
    \Pi_0(t_1,\mu_Q^2)\Big(1+\Pi_0^\mr{(1)}(t_1,\mu_Q^2)\Big)\Big(w_1(\Phi_1)+w_1^{(1)}(\Phi_1)\Big)
    \,\mr{B}_1(\Phi_1)\,\bar{\mc{F}}_1(t_1,O)\\
    &\quad+\int_{q_{T,\mr{cut}}}\bs\done\Phi_1\,
      \Big[1-\Pi_0(t_1,\mu_Q^2)\Big(1+\Pi_0^\mr{(1)}(t_1,\mu_Q^2)\Big)\Big]\,\tilde{\mr{B}}_1^{\rm{R}}(\Phi_1)\,O(\Phi_0)\\
    &\qquad\qquad\qquad\qquad\qquad~
    +\int_{q_{T,\mr{cut}}}\bs\done\Phi_1
    \Pi_0(t_1,\mu_Q^2)\Big(1+\Pi_0^\mr{(1)}(t_1,\mu_Q^2)\Big)\,\tilde{\mr{B}}_1^{\rm{R}}(\Phi_1)\,\bar{\mc{F}}_1(t_1,O)\\
    &\quad+\int_{q_{T,\mr{cut}}}\bs\done\Phi_2\,
      \Big[1-\Pi_0(t_1,\mu_Q^2)\Big(1+\Pi_0^\mr{(1)}(t_1,\mu_Q^2)\Big)\Big]\,\mr{H}_1^{\mr{R}}(\Phi_2)\,O(\Phi_0)\\
    &\qquad\qquad\qquad\qquad\qquad~
    +\int_{q_{T,\mr{cut}}}\bs\done\Phi_2\,
      \Pi_0(t_1,\mu_Q^2)\Big(1+\Pi_0^\mr{(1)}(t_1,\mu_Q^2)\Big)\,\mr{H}_1^{\mr{R}}(\Phi_2)\,\mc{F}_2(t_2,O)\\
    &\quad+\int_{q_{T,\mr{cut}}}\bs\done\Phi_2\,
      \,\mr{H}_1^{\mr{E}}(\Phi_2)\,\mc{F}_2(t_2,O)\;,
  \end{split}
\end{equation}
This again integrates to the NNLO cross section, while preserving the parton
shower accuracy as well as the 1-jet NLO result up to higher orders in $\alpha_s$.
The changes to the one-jet contribution of the \UNNLOPS formula may be interpreted as
a complete subtraction of $\order(\Gamma)$ contributions from the parton shower,
where the branching probability, $\Gamma(t,Q^2)=\done\log\Pi(t,Q^2)/\done t$, 
is the natural expansion parameter of the resummation.
Compared to the procedure described in~\cite{Hoeche:2014aia}, we obtain 
the following additional contribution to the 1-jet bin:
\begin{equation}
  \begin{split}
   \qquad&\int_{q_{T,\mr{cut}}}\bs\done\Phi_1\,
    \Pi_0(t_1,\mu_Q^2)\,
    \Pi_0^\mr{(1)}(t_1,\mu_Q^2)\,
    \left[ \tilde{\mr{B}}_1^{\rm{R}}(\Phi_1)
         + \mr{B}_1(\Phi_1)\,
           \left(w_1(\Phi_1) - 1 + w_1^{(1)}(\Phi_1)\right)\right]
    \bar{\mc{F}}_1(t_1,O)\\
  +&\int_{q_{T,\mr{cut}}}\bs\done\Phi_2\,
    \Pi_0(t_1,\mu_Q^2)\,
    \Pi_0^\mr{(1)}(t_1,\mu_Q^2)\,
    \mr{H}_1^{\mr{R}}(\Phi_2)\,\mc{F}_2(t_2,O)
  \end{split}
\end{equation}
These terms are clearly beyond the formal accuracy of the \UNNLOPS method. 
Including them emphasizes the fixed-order result at medium $q_T$ and therefore
removes some arbitrariness introduced by the parton shower resummation.
It can thus be expected to be an improvement over the method presented 
in~\cite{Hoeche:2014aia}, even though a thorough assessment can only be
made after matching to N$^3$LO fixed-order results. The implementation of 
Eq.~\eqref{eq:nnlo_ps_revised} in a Monte-Carlo event generator is described
in Appendix~\ref{sec:steps}.

In Eq.~\eqref{eq:nnlo_ps_revised}, $\bar{\bar{\mr{B}}}_0^{q_{T,\rm cut}}$ does not affect exclusive observables.
However, the hard function coming from the square of Wilson coefficients of the Higgs effective theory in $\bar{\bar{\mr{B}}}_0$ are universal,
and they should in fact appear also in differential distributions at higher orders, as they factorize 
from real-radiative corrections. It might be beneficial to single out such contributions. 
We therefore define two variants of the \UNNLOPS method, which use the Wilson coefficients
to improve the simulation of the one-jet process. This is very similar to the \MCatNLO and
\POWHEG methods, as discussed in~\cite{Hoeche:2014aia}.

The two possible ways of dealing with the factorized hard function $H(Q^2,\mu^2)=\sum [\alpha_s(\mu^2)/(4\pi)]^n h^{(n)}(Q^2,\mu^2)$ are: 
\begin{itemize}
\item `{\it individual}' matching
  \begin{itemize}
  \item Terms multiplying $h^{(0)}$ are matched using \UNNLOPS
  \item Terms multiplying $h^{(1)}$ are matched using \MCatNLO
  \item Terms multiplying $h^{(2)}$ are showered
  \end{itemize}
\item `{\it factorized}' matching
  \begin{itemize}
  \item The NNLO result in HEFT, ignoring the Wilson coefficient, is matched using \UNNLOPS 
  \item The matched result is multiplied by $H$ in Eq.~\eqref{eq:hard_function}
  \end{itemize}
\end{itemize}
Note that the factorized matching increases the cross section by a few percent 
(see Sec.~\ref{sec:results}). This increase can legitimately be considered as part
of the large NNLO theoretical uncertainty in the Higgs-production process.

\section{Results}
\label{sec:results}
\begin{table}
  \begin{center}
  \begin{tabular}{l@{\hspace*{5mm}}r@{}l@{}l@{\hspace*{5mm}}r@{}l@{}l
      @{\hspace*{5mm}}r@{}l@{}l@{\hspace*{5mm}}r@{}l@{}l}\hline
    $E_{\rm cms}$ & \multicolumn{3}{c}{7 TeV} & \multicolumn{3}{c}{14 TeV} &
    \multicolumn{3}{c}{33 TeV} & \multicolumn{3}{c}{100 TeV} \\\hline\hline
    HNNLO \vphantom{$\int^{^A}_{_B}$} &
    13.&494(7)&$^{+1.436}_{-1.382}$ pb&
    44.&550(16)&$^{+4.293}_{-3.954}$ pb&
    160.&84(13)&$^{+13.29}_{-12.36}$ pb&
    --&&\\
    SHERPA &
    13.&515(7)&$^{+1.443}_{-1.382}$ pb&
    44.&559(36)&$^{+4.226}_{-3.929}$ pb&
    160.&39(17)&$^{+13.47}_{-11.88}$ pb&
    670.&1(10)&$^{+47.9}_{-39.4}$ pb\\[1mm]
    \hline
  \end{tabular}
  \caption{Total cross sections at varying center-of-mass energy 
    for a $pp$-collider. Uncertainties from scale variations are given 
    as sub-/superscripts. Statistical uncertainties from Monte-Carlo 
    integration are quoted in parentheses.
    \label{tab:xs_comparison}}
  \end{center}
\end{table}
\begin{figure}
  \begin{center}
    \includegraphics[width=0.495\textwidth]{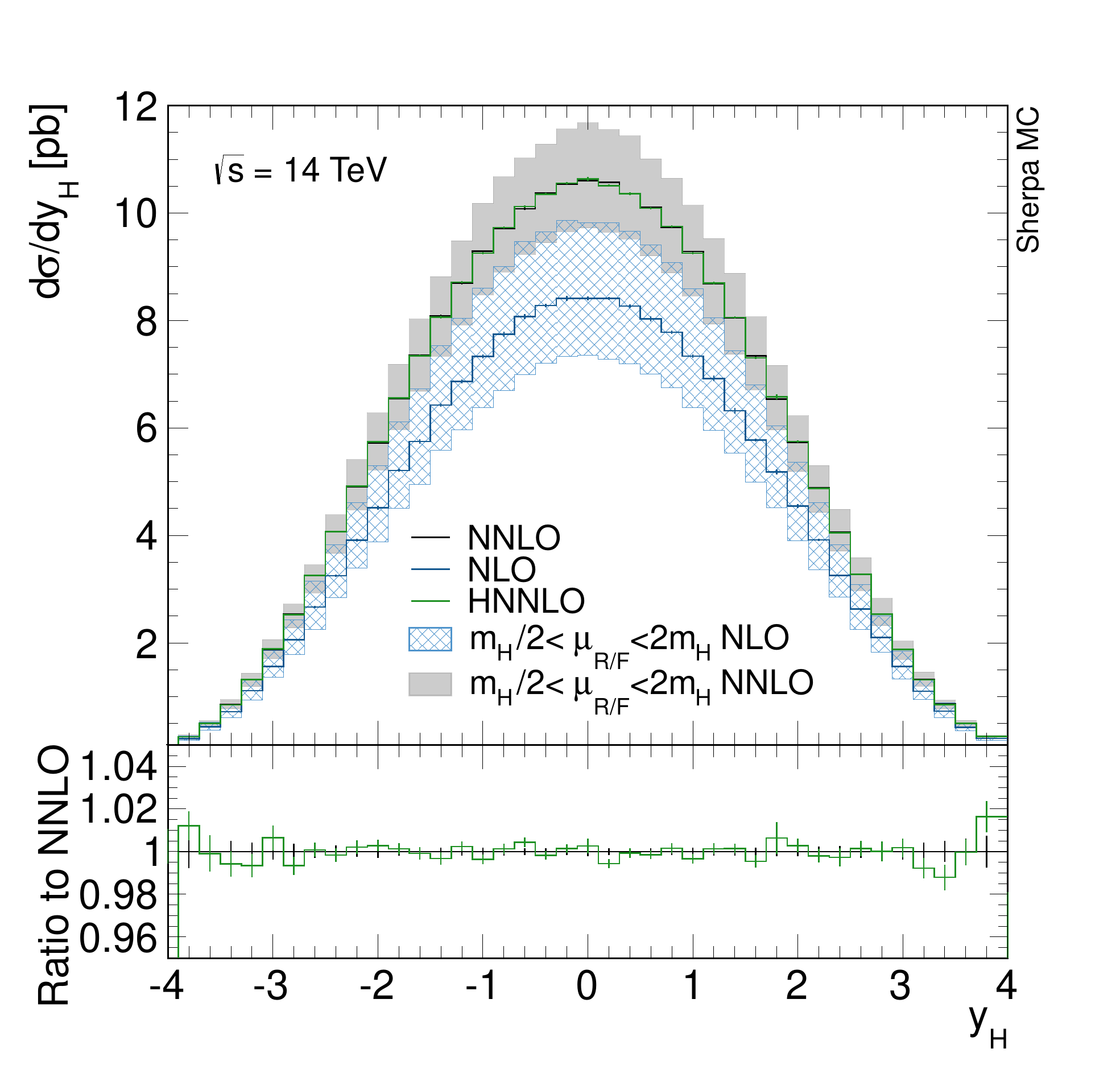}\hfill
    \includegraphics[width=0.495\textwidth]{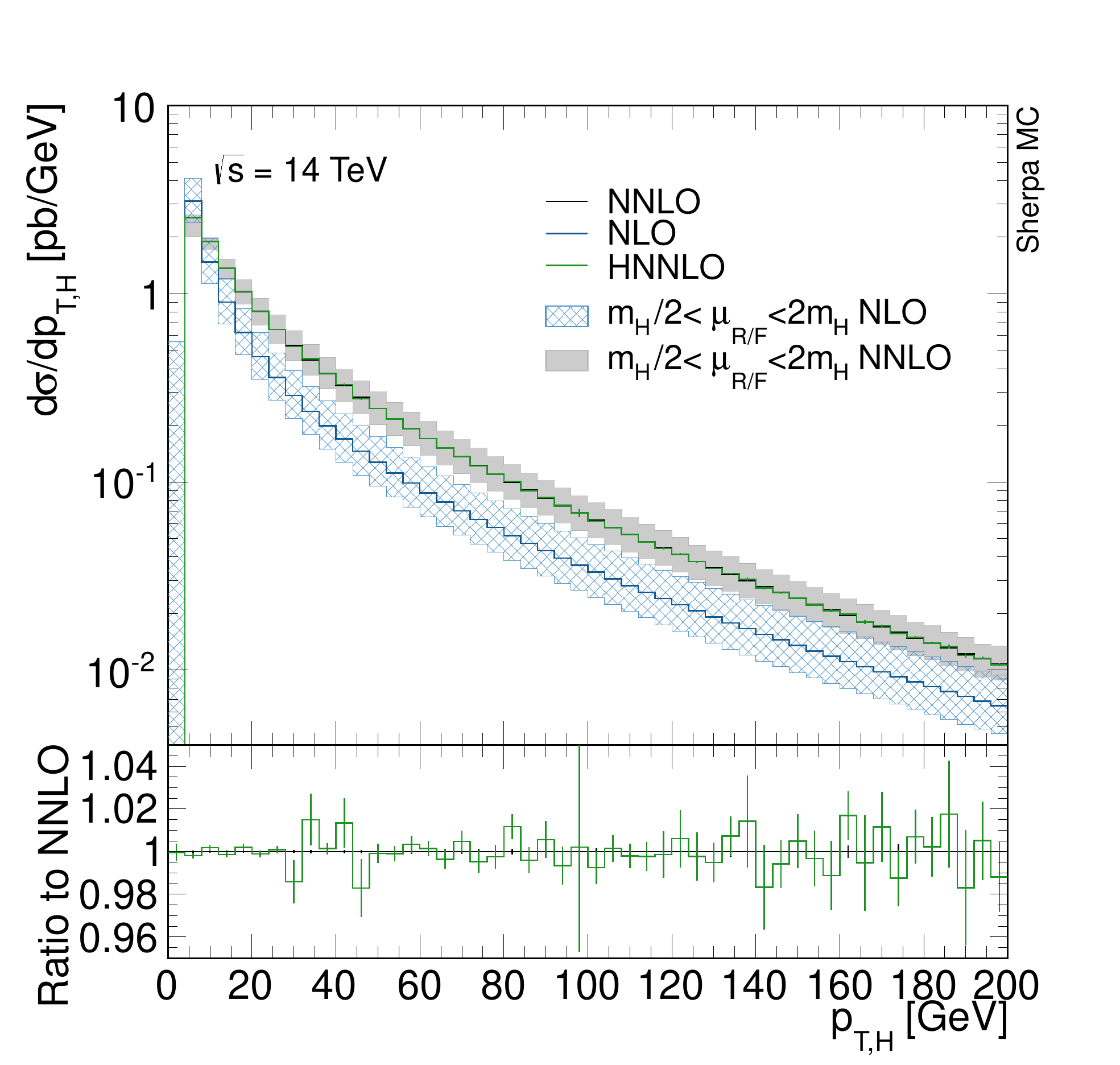}
  \end{center}
  \caption{Rapidity spectrum (left) and transverse momentum spectrum (right)
    of the Higgs boson, computed at fixed order and compared between Sherpa
    and \HNNLO.
    \label{fig:h_pt}}
\end{figure}
This section presents results using an implementation of the
\UNNLOPS algorithm in the event generator Sherpa~\cite{Gleisberg:2003xi,*Gleisberg:2008ta}.
We use a parton shower~\cite{Schumann:2007mg} based on Catani-Seymour
dipole subtraction~\cite{Catani:1996vz,*Catani:2002hc}. NLO virtual corrections for the
one-jet process~\cite{Ravindran:2002dc} are taken from \MCFM~\cite{Campbell:2010ff,*MCFM}. 
Dipole subtraction is performed using Amegic~\cite{Krauss:2001iv,*Gleisberg:2007md}
and cross-checked with Comix~\cite{Gleisberg:2008fv}. We use the MSTW 2008 
PDF set~\cite{Martin:2009iq} and the corresponding definition of the 
running coupling. We work in the five flavor scheme. Electroweak parameters are given 
as $G_F=1.1663787\cdot 10^{-5}\;{\rm GeV}^{-2}$, $m_H=125\;{\rm GeV}$.
The results are derived in the limit $m_t\gg m_H$. 
Predictions for finite $m_t$ will be given elsewhere.

In order to cross-check our implementation we first compare the total cross section
to results obtained from \HNNLO~\cite{Catani:2007vq,Grazzini:2008tf,*Catani:2008me}.
Table~\ref{tab:xs_comparison} shows that the predictions
agree within the permille-level statistical uncertainty of the 
Monte-Carlo integration. Additionally, we have checked that our 
results are identical when varying $q_{T,\rm cut}$ between 
0.1~GeV and 1~GeV. The default value is $q_{T,\rm cut}=$1~GeV.
Figure~\ref{fig:h_pt} shows a comparison of the Higgs rapidity and transverse 
momentum spectrum between Sherpa and \HNNLO. The excellent agreement over a wide
range of phase space confirms the correct implementation of the NNLO calculation
in Sherpa.

\begin{figure}
  \begin{center}
    \includegraphics[width=0.495\textwidth]{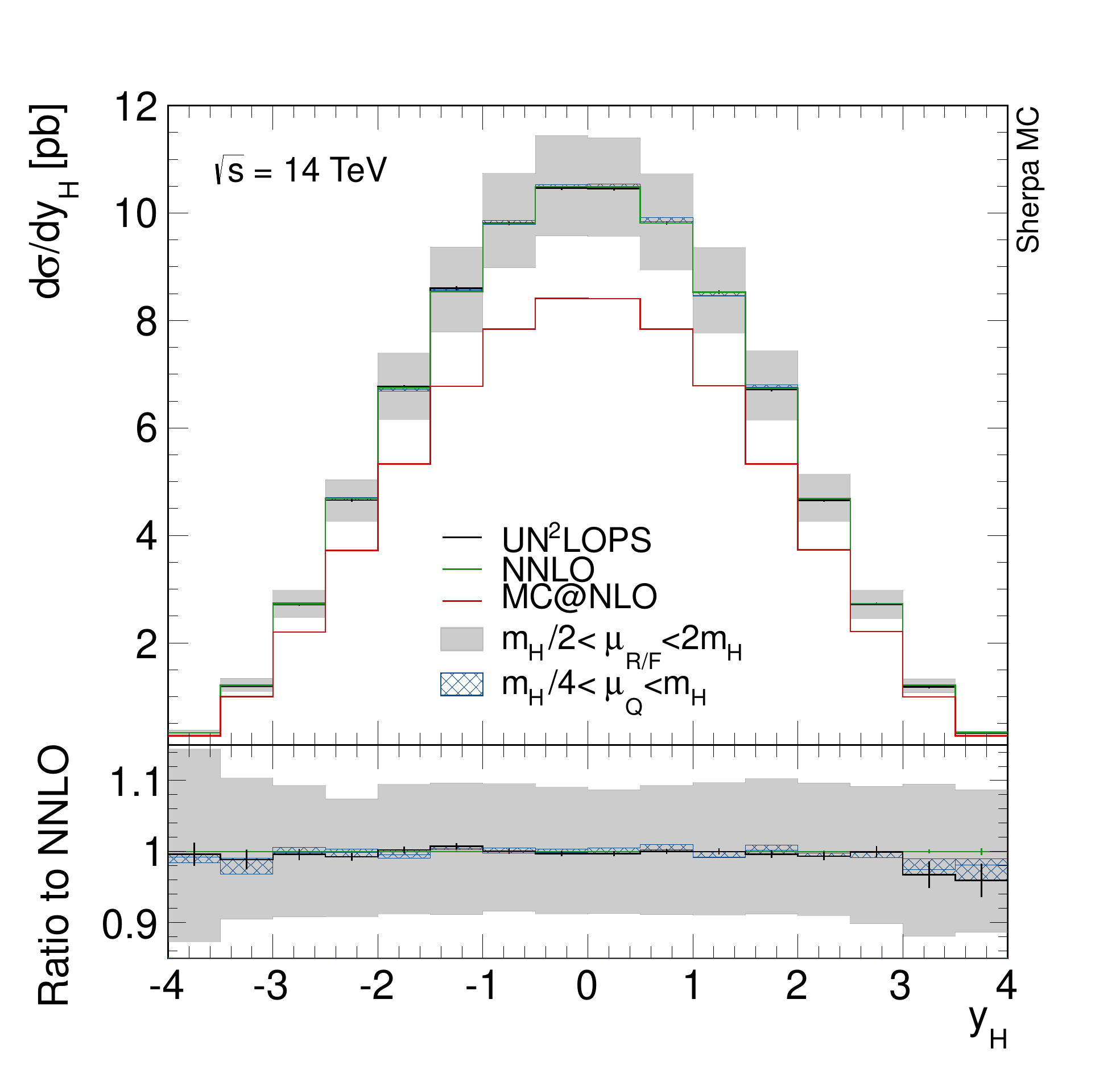}\hfill
    \includegraphics[width=0.495\textwidth]{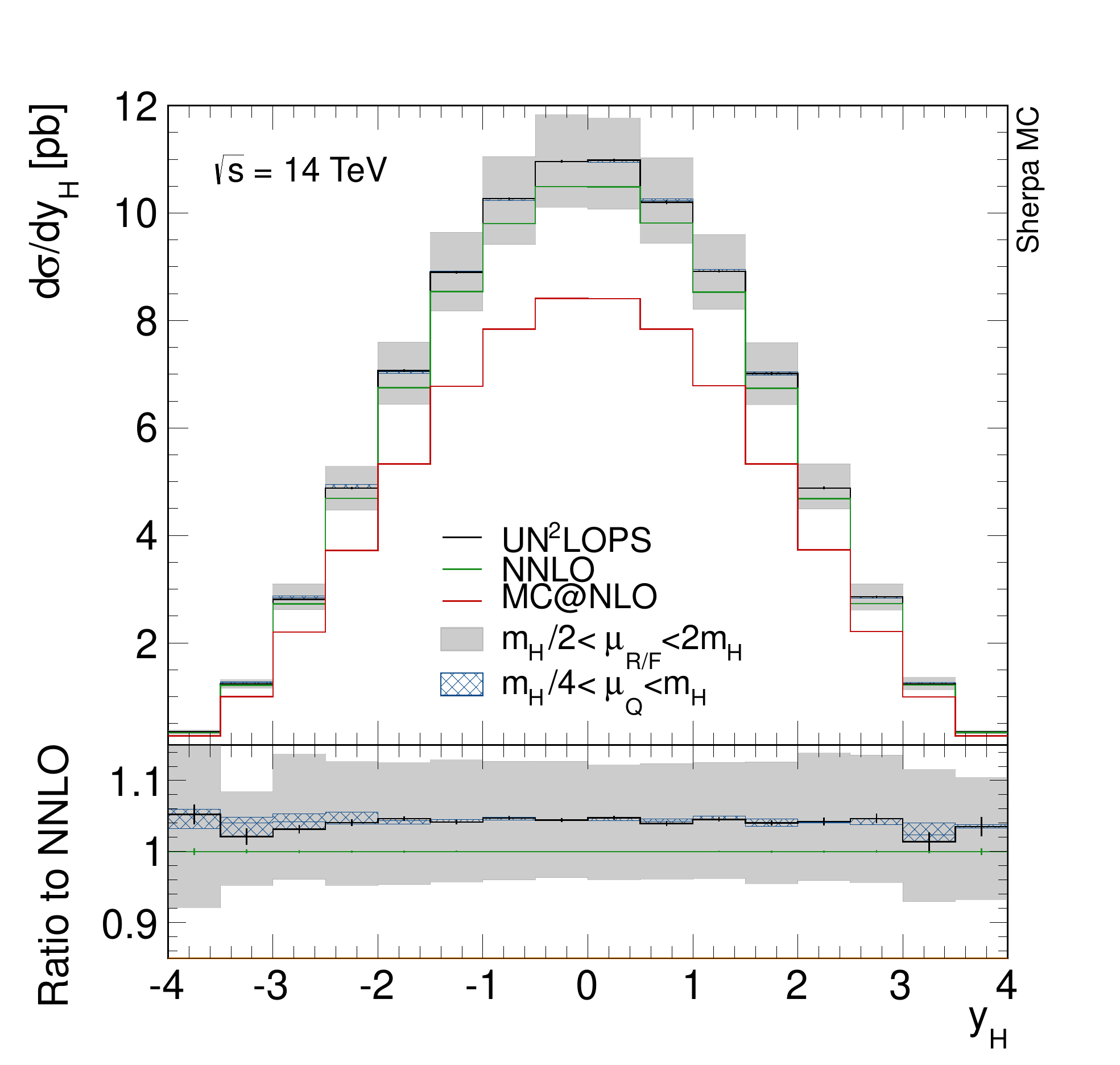}
  \end{center}
  \caption{Rapidity spectrum of the Higgs boson in 
    individual matching (left) and factorized matching (right).
    See Sec.~\ref{sec:higgs} for details.
    \label{fig:h_y_matched}}
\end{figure}
\begin{figure}
  \begin{center}
    \includegraphics[width=0.495\textwidth]{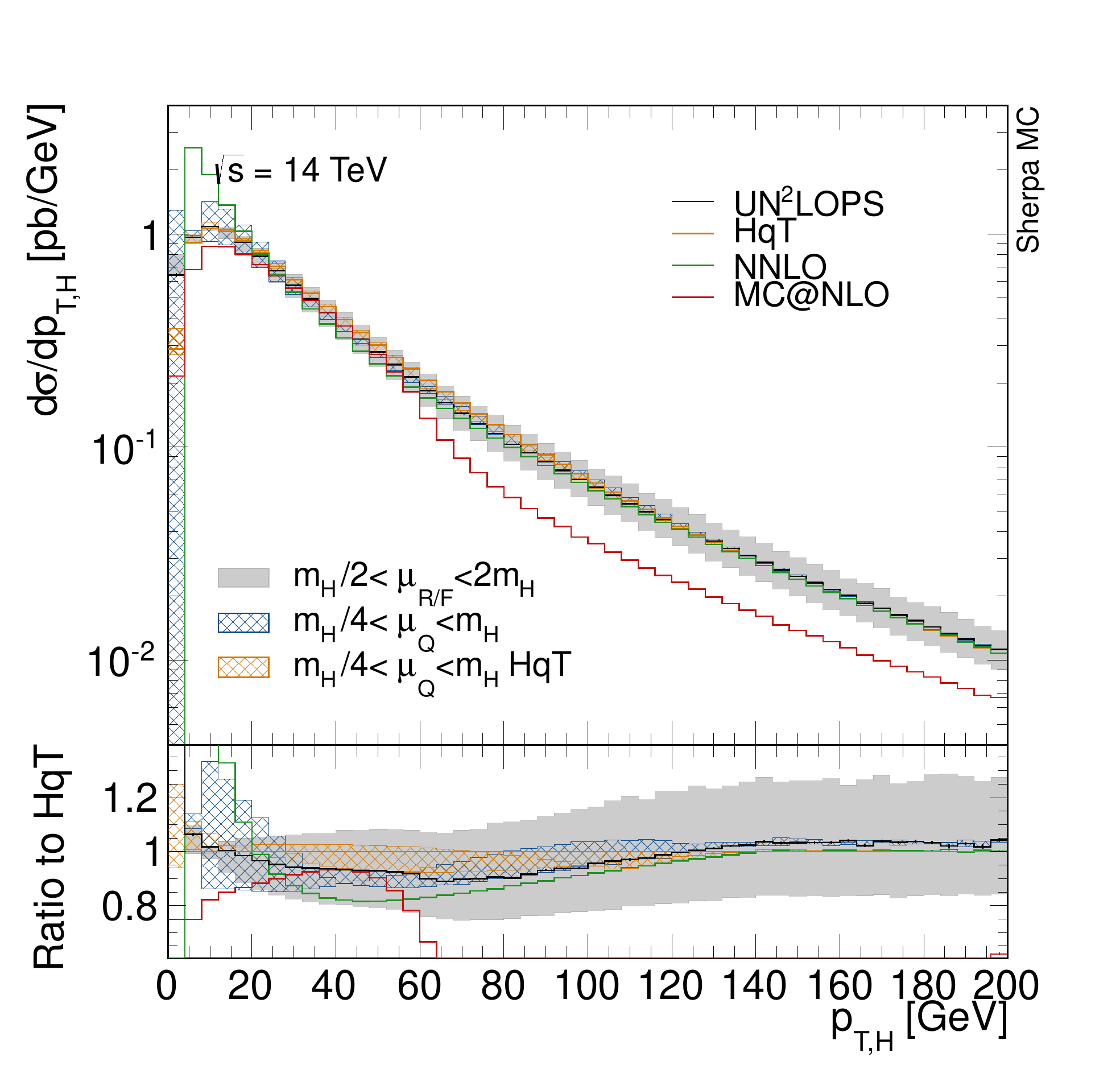}\hfill
    \includegraphics[width=0.495\textwidth]{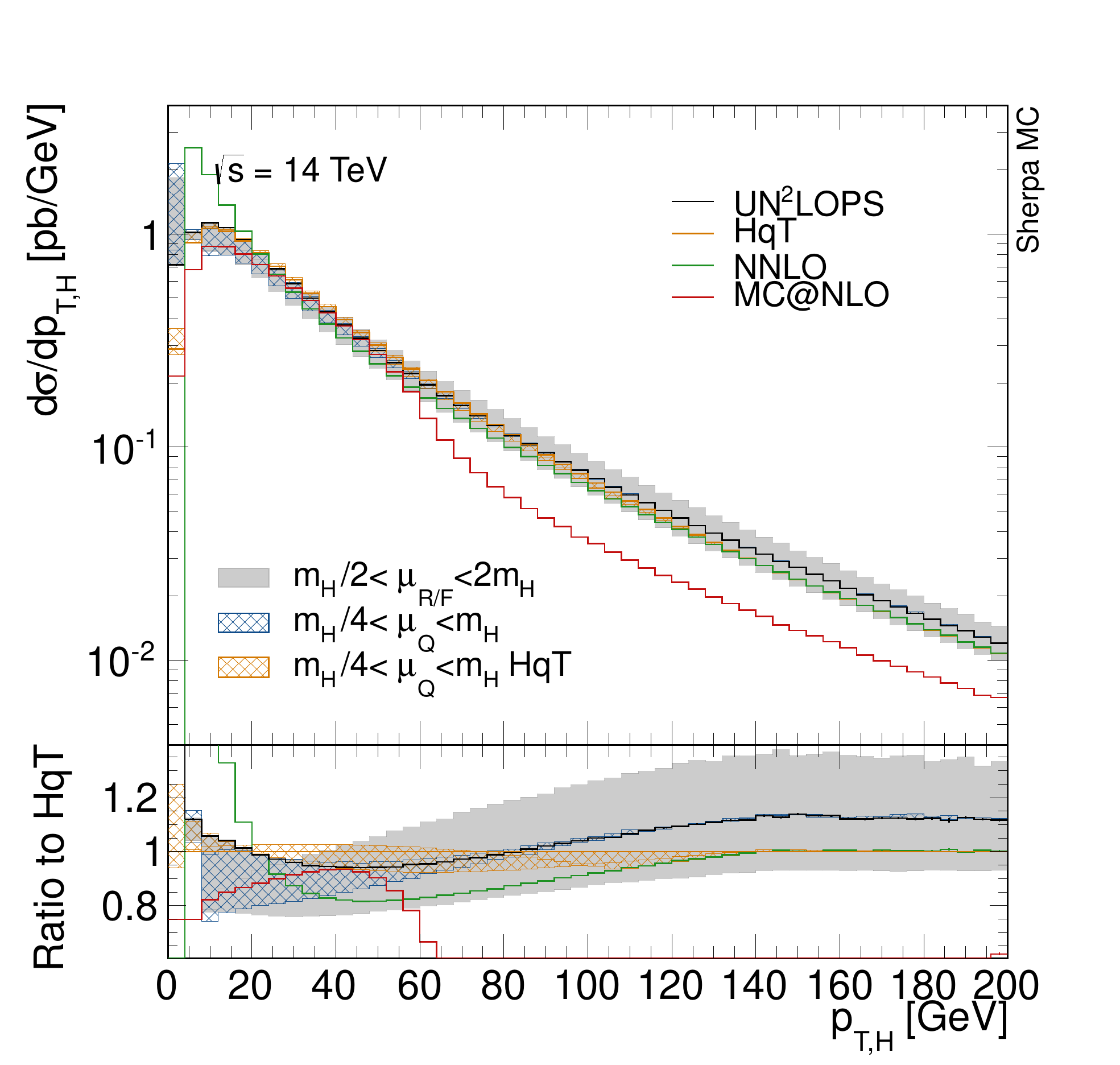}
  \end{center}
  \caption{Transverse momentum spectrum of the Higgs boson in
    individual matching (left) and factorized matching (right).
    See Sec.~\ref{sec:higgs} for details.
    \label{fig:h_pt_matched}}
\end{figure}
Figure~\ref{fig:h_y_matched} compares fixed-order predictions for the 
rapidity spectrum of the Higgs boson to matched results from \UNNLOPS.
Both the individual and factorized matching approach, introduced in Sec.~\ref{sec:higgs}
yield perfect agreement for the shape of the distribution, while the factorized
matching also slightly increases the cross section (see Sec.~\ref{sec:higgs}).

Figure~\ref{fig:h_pt_matched} compares the \UNNLOPS matched results for the 
Higgs boson transverse momentum to predictions from 
HqT~\cite{Bozzi:2003jy,*Bozzi:2005wk,*deFlorian:2011xf},
which performs an analytic matching of the $q_T$ spectrum at NLO+NNLL accuracy.
As expected, the resummation uncertainty in \UNNLOPS is larger. Nevertheless,
the central predictions agree quite well. This indicates that the impact of 
possible higher logarithmic contributions should be small enough to be neglected 
for the purpose of event generation at the 14~TeV LHC, provided that the
central resummation scale is set appropriately. Similar findings were reported
for $t\bar{t}$ production in~\cite{Corke:2010zj}.

The zero-$q_T$ bin is clearly problematic. This can be understood as follows:
In our calculation the $q_T$ spectrum is described only at NLO+NLL 
accuracy~\cite{Catani:1990rr,Platzer:2009jq}. Therefore it suffers from 
large scale variations, particularly in the soft and collinear region.
Equation~\ref{eq:nnlo_ps_revised} implies that an increased veto probability
in this region also increases the cross section in the zero-$q_T$ bin.
The large variation at zero-$q_T$ should thus be reduced significantly
once the parton shower is amended with higher-logarithmic resummation.
Note, however, that no such variation is present for inclusive observables
like the Higgs-boson rapidity spectrum, Fig.~\ref{fig:h_y_matched}.

\section{Conclusions}
\label{sec:conclusions}
We presented the first application of the \UNNLOPS matching procedure to Higgs-boson
production through gluon fusion. This reaction suffers from large higher-order corrections,
and several refinements of the original \UNNLOPS approach are suggested to improve the matching.
They allow to obtain phenomenologically useful results despite the low logarithmic accuracy 
of the parton shower compared to analytic approaches. Our predictions are in fair agreement 
with higher logarithmic resummation for a resummation scale of $\mu_Q\sim m_H/2$.

We also provide an independent implementation of a fully differential NNLO
calculation of Higgs-boson production at hadron colliders, using the $q_T$-cutoff
method, which allows the production of LHEF files~\cite{Alwall:2006yp,*Butterworth:2014efa}
or NTuple files~\cite{Brun:1997pa,*Bern:2013zja} containing NNLO event information at parton level.

Due to the fully exclusive nature of our simulations, it is straightforward to combine 
them with higher-multiplicity NLO calculations using an extension of the \MENLOPS 
method~\cite{Hamilton:2010wh,*Hoeche:2010kg} to NNLO. This can be achieved by separating 
the generating functionals into a hard and a soft part and using appropriately
weighted NLO calculations in the hard jet domain. A similar scheme, which also
preserves the total cross section, could be based on the \UNLOPS method~\cite{Lonnblad:2012ix,Platzer:2012bs}.
Such a simulation will improve upon our predictions as soon as the Higgs-boson 
plus two-jet process is included at NLO. We leave the detailed study 
of the phenomenological implications to future work.

\section*{Acknowledgments}
We thank Marek Sch{\"o}nherr and Thomas L{\"u}bbert for discussions and Valentin Hirschi 
and Gionata Luisoni for careful cross-checks of the virtual corrections to Higgs plus 
one-jet production. We are indebted to Lance Dixon, Frank Krauss, Leif L{\"o}nnblad 
and HuaXing Zhu for helpful conversations and comments on the manuscript. 
This work was supported by the US Department of Energy under contract DE--AC02--76SF00515.
We used resources of the National Energy Research Scientific Computing Center, 
which is supported by the Office of Science of the U.S.\ Department of Energy 
under Contract No.\ DE--AC02--05CH11231.

\begin{appendix}
\section{\UNNLOPS step-by-step}
\label{sec:steps}

This appendix provides a step-by-step guide to the \UNNLOPS method,
which allows to implement the technique in other processes of interest.
We focus on reactions at hadron colliders, the generalization to lepton
colliders being a straightforward modification. We assume that a vetoed
NNLO cross section, computed along the lines of Sec.~\ref{sec:nnlo} exists.
Any other method to provide this cross section can be used, for example 
a fully exclusive NNLO calculation restricted to $q_T<q_{T,\rm cut}$ 
by means of a $q_T$ veto.
\begin{enumerate}
\item Generate a parton-level event according to $\bar{\bar{\mr{B}}}_0^{q_{T,\rm cut}}$,
  $\mr{B}_1$, $\tilde{\mr{B}}_1^{\rm R}$ or $\mr{H}_1$.
\item If $\bar{\bar{\mr{B}}}_0^{q_{T,\rm cut}}$ is selected, skip the parton shower step.
\item\label{step:cluster}
  If $\mr{B}_1$, $\tilde{\mr{B}}_1$ of $\mr{H}_1$ is selected, apply
  the clustering algorithm described in~\cite{Andre:1997vh} to define 
  a parton-shower history. If no history is found on $\mr{H}_1$ events,
  they are classified as exceptional. In this case, simply run the 
  parton shower starting from the two-particle state.
\item In $\mr{B}_1$ events, reweight with $w_1$ in Eq.~\eqref{eq:unlops_weight}
  and subtract the first-order expansion, $w_1^{(1)}$.
\item Run a truncated parton shower on the zero-jet configuration as identified
  by backward clustering in step~\ref{step:cluster}.
  Skip the first emission~\cite{Hoeche:2012yf}.
  If the parton shower generates a second emission, reduce the one- or two-particle
  configuration to the zero-particle configuration identified in step~\ref{step:cluster}
  and do not apply any further parton shower or \MCatNLO.
  If the parton shower does not generate a second emission, and the event
  was of $\tilde{\mr{B}}_1$ type, perform the \MCatNLO starting from the
  one-particle state. If the event was of $\mr{H}_1^{\rm R}$ type, 
  run the parton shower, starting from the two-particle state.
\end{enumerate}

\end{appendix}

\bibliography{journal}

\end{document}